
\newif\iflanl
\openin 1 lanlmac
\ifeof 1 \lanlfalse \else \lanltrue \fi
\closein 1
\iflanl
    \input lanlmac
\else
    \message{[lanlmac not found - use harvmac instead}
    \input harvmac
\fi
\newif\ifhypertex
\ifx\hyperdef\UnDeFiNeD
    \hypertexfalse
    \message{[HYPERTEX MODE OFF}
    \def\hyperref#1#2#3#4{#4}
    \def\hyperdef#1#2#3#4{#4}
    \def\hypernoname{}
    \def\e@tf@ur#1{}
    \def\hth/#1#2#3#4#5#6#7{{\tt hep-th/#1#2#3#4#5#6#7}}
    \def\CERN{\address{CERN, Geneva, Switzerland}}
\else
    \hypertextrue
    \message{[HYPERTEX MODE ON}
  \def\hth/#1#2#3#4#5#6#7{
  {\tt hep-th/#1#2#3#4#5#6#7}}
\def\CERN{\address{

Theory Division, CERN, Geneva, Switzerland}}
\fi
\newif\ifdraft

\noblackbox
\catcode`\@=11
\newif\iffrontpage
\ifx\answ\bigans
\def\titleft{\titsm}
\magnification=1200\baselineskip=14pt plus 2pt minus 1pt
%
\advance\hoffset by-0.075truein
\advance\voffset by1.truecm
\hsize=6.15truein\vsize=600.truept\hsbody=\hsize\hstitle=\hsize
\else\let\lr=L
\def\titleft{\titla}
\magnification=1000\baselineskip=14pt plus 2pt minus 1pt
%
\hoffset=-0.75truein\voffset=-.0truein
\vsize=6.5truein
\hstitle=8.truein\hsbody=4.75truein
\fullhsize=10truein\hsize=\hsbody
\fi
\parskip=4pt plus 15pt minus 1pt
%
\newif\iffigureexists
\newif\ifepsfloaded
\def\epsfcheck{
\ifdraft
\input epsf\epsfloadedtrue
\else
  \openin 1 epsf
  \ifeof 1 \epsfloadedfalse \else \epsfloadedtrue \fi
  \closein 1
  \ifepsfloaded
    \input epsf
  \else
\immediate\write20{NO EPSF FILE --- FIGURES WILL BE IGNORED}
  \fi
\fi
\def\epsfcheck{}}
\def\checkex#1{
\ifdraft
\figureexistsfalse\immediate%
\write20{Draftmode: figure #1 not included}
\else\relax
    \ifepsfloaded \openin 1 #1
        \ifeof 1
           \figureexistsfalse
  \immediate\write20{FIGURE FILE #1 NOT FOUND}
        \else \figureexiststrue
        \fi \closein 1
    \else \figureexistsfalse
    \fi
\fi}
\def\missbox#1#2{$\vcenter{\hrule
\hbox{\vrule height#1\kern1.truein
\raise.5truein\hbox{#2} \kern1.truein \vrule} \hrule}$}
\def\lfig#1{
\let\labelflag=#1%
\def\numb@rone{#1}%
\ifx\labelflag\UnDeFiNeD%
{\xdef#1{\the\figno}%
\writedef{#1\leftbracket{\the\figno}}%
\global\advance\figno by1%
}\fi{\hyperref{}{figure}{{\numb@rone}}{Fig.{\numb@rone}}}}
\def\figinsert#1#2#3#4{
\epsfcheck\checkex{#4}%
\def\figsize{#3}%
\let\flag=#1\ifx\flag\UnDeFiNeD
{\xdef#1{\the\figno}%
\writedef{#1\leftbracket{\the\figno}}%
\global\advance\figno by1%
}\fi
\goodbreak\midinsert%
\iffigureexists
\centerline{\epsfysize\figsize\epsfbox{#4}}%
\else%
\vskip.05truein
  \ifepsfloaded
  \ifdraft
  \centerline{\missbox\figsize{Draftmode: #4 not included}}%
  \else
  \centerline{\missbox\figsize{#4 not found}}
  \fi
  \else
  \centerline{\missbox\figsize{epsf.tex not found}}
  \fi
\vskip.05truein
\fi%
{\smallskip%
\leftskip 4pc \rightskip 4pc%
\noindent\ninepoint\sl \baselineskip=11pt%
{\bf{\hyperdef\hypernoname{figure}{{#1}}{Fig.{#1}}}:~}#2%
\smallskip}\bigskip\endinsert%
}
%
\font\bigit=cmti10 scaled \magstep1

\font\titla=cmr10 scaled\magstep3
\font\tenmss=cmss10
\font\absmss=cmss10 scaled\magstep1

\newfam\mssfam
\font\footrm=cmr8  \font\footrms=cmr5
\font\footrmss=cmr5   \font\footi=cmmi8
\font\footis=cmmi5   \font\footiss=cmmi5
\font\footsy=cmsy8   \font\footsys=cmsy5
\font\footsyss=cmsy5   \font\footbf=cmbx8
\font\footmss=cmss8
\def\footfont{\def\rm{\fam0\footrm}
\textfont0=\footrm \scriptfont0=\footrms
\scriptscriptfont0=\footrmss
\textfont1=\footi \scriptfont1=\footis
\scriptscriptfont1=\footiss
\textfont2=\footsy \scriptfont2=\footsys
\scriptscriptfont2=\footsyss
\textfont\itfam=\footi \def\it{\fam\itfam\footi}
\textfont\mssfam=\footmss \def\mss{\fam\mssfam\footmss}
\textfont\bffam=\footbf \def\bf{\fam\bffam\footbf} \rm}
\def\tenpoint{\def\rm{\fam0\tenrm}
\textfont0=\tenrm \scriptfont0=\sevenrm
\scriptscriptfont0=\fiverm
\textfont1=\teni  \scriptfont1=\seveni
\scriptscriptfont1=\fivei
\textfont2=\tensy \scriptfont2=\sevensy
\scriptscriptfont2=\fivesy
\textfont\itfam=\tenit \def\it{\fam\itfam\tenit}
\textfont\mssfam=\tenmss \def\mss{\fam\mssfam\tenmss}
\textfont\bffam=\tenbf \def\bf{\fam\bffam\tenbf} \rm}
\ifx\answ\bigans\def\abstractfont{\tenpoint}\else
\def\abstractfont{\def\rm{\fam0\absrm}
\textfont0=\absrm \scriptfont0=\absrms
\scriptscriptfont0=\absrmss
\textfont1=\absi \scriptfont1=\absis
\scriptscriptfont1=\absiss
\textfont2=\abssy \scriptfont2=\abssys
\scriptscriptfont2=\abssyss
\textfont\itfam=\bigit \def\it{\fam\itfam\bigit}
\textfont\mssfam=\absmss \def\mss{\fam\mssfam\absmss}
\textfont\bffam=\absbf \def\bf{\fam\bffam\absbf}\rm}\fi
%
\def\f@@t{\baselineskip10pt\lineskip0pt\lineskiplimit0pt
\bgroup\aftergroup\@foot\let\next}
\setbox\strutbox=\hbox{\vrule height 8.pt depth 3.5pt width\z@}
\def\vfootnote#1{\insert\footins\bgroup
\baselineskip10pt\footfont
\interlinepenalty=\interfootnotelinepenalty
\floatingpenalty=20000
\splittopskip=\ht\strutbox \boxmaxdepth=\dp\strutbox
\leftskip=24pt \rightskip=\z@skip
\parindent=12pt \parfillskip=0pt plus 1fil
\spaceskip=\z@skip \xspaceskip=\z@skip
\Textindent{$#1$}\footstrut\futurelet\next\fo@t}
\def\Textindent#1{\noindent\llap{#1\enspace}\ignorespaces}
\def\foot{\global\advance\ftno by1%
\attach{\hyperref{}{footnote}{\the\ftno}{\footsymbolgen}}%
\vfootnote{\hyperdef\hypernoname{footnote}{\the\ftno}{\footsymbol}}}%
\def\footnote#1{\global\advance\ftno by1%
\attach{\hyperref{}{footnote}{\the\ftno}{#1}}%
\vfootnote{\hyperdef\hypernoname{footnote}{\the\ftno}{#1}}}%
\newcount\lastf@@t           \lastf@@t=-1
\newcount\footsymbolcount    \footsymbolcount=0
\global\newcount\ftno \global\ftno=0
\def\footsymbolgen{\relax\footsym
\global\lastf@@t=\pageno\footsymbol}
\def\footsym{\ifnum\footsymbolcount<0
\global\footsymbolcount=0\fi
{\iffrontpage \else \advance\lastf@@t by 1 \fi
\ifnum\lastf@@t<\pageno \global\footsymbolcount=0
\else \global\advance\footsymbolcount by 1 \fi }
\ifcase\footsymbolcount
\fd@f\dagger\or \fd@f\diamond\or \fd@f\ddagger\or
\fd@f\natural\or \fd@f\ast\or \fd@f\bullet\or
\fd@f\star\or \fd@f\nabla\else \fd@f\dagger
\global\footsymbolcount=0 \fi }
\def\fd@f#1{\xdef\footsymbol{#1}}
\def\space@ver#1{\let\@sf=\empty \ifmmode #1\else \ifhmode
\edef\@sf{\spacefactor=\the\spacefactor}
\unskip${}#1$\relax\fi\fi}
\def\attach#1{\space@ver{\strut^{\mkern 2mu #1}}\@sf}
%
\newif\ifnref
\def\rrr#1#2{\relax\ifnref\nref#1{#2}\else\ref#1{#2}\fi}
\def\ldf#1#2{\begingroup\obeylines
\gdef#1{\rrr{#1}{#2}}\endgroup\unskip}

\def\doubref#1#2{\refs{{#1},{#2}}}

\nreffalse
\def\refout{\listrefs}
%
\def\eqn#1{\xdef #1{(\noexpand\hyperref{}%
{equation}{\secsym\the\meqno}%
{\secsym\the\meqno})}\eqno(\hyperdef\hypernoname{equation}%
{\secsym\the\meqno}{\secsym\the\meqno})\eqlabeL#1%
\writedef{#1\leftbracket#1}\global\advance\meqno by1}
\def\eqnalign#1{\xdef #1{\noexpand\hyperref{}{equation}%
{\secsym\the\meqno}{(\secsym\the\meqno)}}%
\writedef{#1\leftbracket#1}%
\hyperdef\hypernoname{equation}%
{\secsym\the\meqno}{\e@tf@ur#1}\eqlabeL{#1}%
\global\advance\meqno by1}
\def\eqnalign#1{\xdef #1{(\secsym\the\meqno)}
\writedef{#1\leftbracket#1}%
\global\advance\meqno by1 #1\eqlabeL{#1}}
%

%
\def\chap#1{\newsec{#1}}
\def\chapter#1{\chap{#1}}
\def\sect#1{\subsec{#1}}
\def\section#1{\sect{#1}}
\def\\{\ifnum\lastpenalty=-10000\relax
\else\hfil\penalty-10000\fi\ignorespaces}
\def\note#1{\leavevmode%
\edef\@@marginsf{\spacefactor=\the\spacefactor\relax}%
\ifdraft\strut\vadjust{%
\hbox to0pt{\hskip\hsize%
\ifx\answ\bigans\hskip.1in\else\hskip .1in\fi%
\vbox to0pt{\vskip-\dp
\strutbox\sevenbf\baselineskip=8pt plus 1pt minus 1pt%
\ifx\answ\bigans\hsize=.7in\else\hsize=.35in\fi%
\tolerance=5000 \hbadness=5000%
\leftskip=0pt \rightskip=0pt \everypar={}%
\raggedright\parskip=0pt \parindent=0pt%
\vskip-\ht\strutbox\noindent\strut#1\par%
\vss}\hss}}\fi\@@marginsf\kern-.01cm}
\def\titlepage{%
\frontpagetrue\nopagenumbers\abstractfont%
\hsize=\hstitle\rightline{\vbox{\baselineskip=10pt%
{\abstractfont\pubnum}}}\pageno=0}
\frontpagefalse
\def\pubnum{}
\def\pdate{\number\month/\number\yearltd}
\def\makefootline{\iffrontpage\vskip .27truein
\line{\the\footline}
\vskip -.1truein\leftline{\vbox{\baselineskip=10pt%
{\abstractfont\pdate}}}
\else\vskip.5cm\line{\hss \tenrm $-$ \folio\ $-$ \hss}\fi}
\def\title#1{\vskip .7truecm\titlestyle{\titleft #1}}
\def\titlestyle#1{\par\begingroup \interlinepenalty=9999
\leftskip=0.02\hsize plus 0.23\hsize minus 0.02\hsize
\rightskip=\leftskip \parfillskip=0pt
\hyphenpenalty=9000 \exhyphenpenalty=9000
\tolerance=9999 \pretolerance=9000
\spaceskip=0.333em \xspaceskip=0.5em
\noindent #1\par\endgroup }
\def\autskip{\ifx\answ\bigans\vskip.5truecm\else\vskip.1cm\fi}
\def\author#1{\vskip .7in \centerline{#1}}

\def\address#1{\ifx\answ\bigans\vskip.2truecm
\else\vskip.1cm\fi{\it \centerline{#1}}}
\def\abstract#1{
\vskip .5in\vfil\centerline
{\bf Abstract}\penalty1000
{{\smallskip\ifx\answ\bigans\leftskip 2pc \rightskip 2pc
\else\leftskip 5pc \rightskip 5pc\fi
\noindent\abstractfont \baselineskip=12pt
{#1} \smallskip}}
\penalty-1000}
\def\endpage{\tenpoint\supereject\global\hsize=\hsbody%
\frontpagefalse\footline={\hss\tenrm\folio\hss}}
%

\def\bfone{\relax{\rm 1\kern-.35em 1}}
\def\inbar{\vrule height1.5ex width.4pt depth0pt}
\def\IC{\relax\,\hbox{$\inbar\kern-.3em{\mss C}$}}
\def\ID{\relax{\rm I\kern-.18em D}}
\def\IF{\relax{\rm I\kern-.18em F}}
\def\IH{\relax{\rm I\kern-.18em H}}
\def\II{\relax{\rm I\kern-.17em I}}
\def\IN{\relax{\rm I\kern-.18em N}}
\def\IP{\relax{\rm I\kern-.18em P}}
\def\IQ{\relax\,\hbox{$\inbar\kern-.3em{\rm Q}$}}
\def\IR{\relax{\rm I\kern-.18em R}}
\font\cmss=cmss10 \font\cmsss=cmss10 at 7pt
\def\ZZ{\relax\ifmmode\mathchoice
{\hbox{\cmss Z\kern-.4em Z}}{\hbox{\cmss Z\kern-.4em Z}}
{\lower.9pt\hbox{\cmsss Z\kern-.4em Z}}
{\lower1.2pt\hbox{\cmsss Z\kern-.4em Z}}\else{\cmss Z\kern-.4em
Z}\fi}
\def\a{\alpha} \def\b{\beta}

\def\L{\Lambda} 
 
\def\cC{{\cal C}} 
\def\cF{{\cal F}}

\def\cL{{\cal L}}

\def\nup#1({Nucl.\ Phys.\ $\us {B#1}$\ (}
\def\plt#1({Phys.\ Lett.\ $\us  {#1}$\ (}
\def\cmp#1({Comm.\ Math.\ Phys.\ $\us  {#1}$\ (}
\def\prp#1({Phys.\ Rep.\ $\us  {#1}$\ (}
\def\prl#1({Phys.\ Rev.\ Lett.\ $\us  {#1}$\ (}
\def\prv#1({Phys.\ Rev.\ $\us  {#1}$\ (}
\def\mpl#1({Mod.\ Phys.\ Let.\ $\us  {A#1}$\ (}
\def\ijmp#1({Int.\ J.\ Mod.\ Phys.\ $\us{A#1}$\ (}
\def\tit#1|{{\it #1},\ }
%

%

\def\ni{\noindent}

\def\bar{\overline}
\def\us#1{\underline{#1}}

\def\Coe#1.#2.{{#1\over #2}}

\def\coe#1.#2.{\relax{\textstyle {#1 \over #2}}\displaystyle}

\def\shalf{\relax{\textstyle {1 \over 2}}\displaystyle}

\def\to{\rightarrow}
\def\notin{\hbox{{$\in$}\kern-.51em\hbox{/}}}

\def\del{\partial}

\def\nex#1{$N\!=\!#1$}

\catcode`\@=12

\def\a{a_1}
\def\ai{a_i}
\def\b{a_2}
\def\ad{a_{D1}}
\def\bd{a_{D2}}
\def\adi{a_{Di}}
\def\cF{{\cal F}}
\def\wan#1{W_{\!A_{#1}}}
\def\D{\Delta}
\def\cW{{\cal W}}
\def\rc{r^{{\rm class}}}
\def\cC{{\cal C}}

\def\g{\gamma}
%
\ldf\SWa{N.\ Seiberg and E.\ Witten, \nup426(1994) 19, \hth/9407087.}
\ldf\SWb{N.\ Seiberg and E.\ Witten, \nup431(1994) 484,
\hth/9408099.}
\ldf\KLTYa{A.\ Klemm, W.\ Lerche, S.\ Theisen and S.\ Yankielowicz,
{\it Simple Singularities and N=2 Supersymmetric Yang-Mills Theory},
preprint CERN-TH.7495/94, LMU-TPW 94/16; \hth/9411048}
\ldf\AF{P. Argyres and A. Faraggi, {\it The Vacuum Structure and
Spectrum
of N=2 Supersymmetric SU(n) Gauge Theory}, preprint IASSNS-HEP-94/94;
\hth/9411057}
\ldf\LGrefs{E. Martinec, \plt 217B(1989) 431;
C.\ Vafa and N.P.\ Warner, \plt218B  (1989) 51.}
\ldf\KLTY{A.\ Klemm, W.\ Lerche, S.\ Theisen
and S. Yankielowicz, in preparation.}
\ldf\Arn{See e.g., V.\ Arnold, A.\ Gusein-Zade and A.\ Varchenko,
{\it Singularities of Differentiable Maps I, II}, Birkh\"auser 1985.}
\ldf\CDFLLR{A. Ceresole, R. D'Auria and T. Regge, \nup414 (1994) 517;
see also: A.\ Ceresole, R. D'Auria, S. Ferrara, W. Lerche, J. Louis
and T. Regge, {\it Picard-Fuchs Equations, Special Geometry and
Target Space Duality}, preprint CERN-TH.7055/93, POLFIS-TH.09/93, to
be published in ``Essays on Mirror Symmetry, Vol. 2", B. Green and
S.-T. Yau, eds.}
\ldf\thooft{G. `t Hooft, \nup190(1981) 455.}
\ldf\theUltimateSpec{P.\ Townsend, \plt202B (1988) 53; C.\ Hull and
P.\ Townsend, {\it Unity of superstring dualities}, preprint
QMW-94-30, \hth/9410167.}
\def\DVV {\rrr\DVV {R.\ Dijkgraaf, E. Verlinde and H. Verlinde,
\nup352(1991) 59.}}
\ldf\NS{N.\ Seiberg, \nup303(1988) 286.}
\ldf\NSbeta{N.\ Seiberg, \plt206(1988) 75.}
\ldf\SCHIF{
V.\ Novikov, M.\ Schifman, A.\ Vainstein,
M\ Voloshin and V.\ Zakharov,
\nup229(1983) 394;
V.\ Novikov, M.\ Schifman, A.\ Vainstein and V.\ Zakharov,
\nup229(1983) 381, 407;
M.\ Schifman, A.\ Vainstein and V.\ Zakharov,
\plt166(1986) 329.}
%
\def\pubnum{
\hbox{CERN-TH.7538/94}
\hbox{LMU-TPW 94/22}
\hbox{hep-th/9412158}}
\def\pdate{}
\titlepage
\title
{On the Monodromies of N=2 Supersymmetric Yang-Mills Theory}
\vskip-1.1cm\autskip
\author{A.\ Klemm, W.\ Lerche, S.\ Yankielowicz\footnote {a}
{On leave of absence from the School of Physics, Raymond and Beverly
Sackler Faculty of Exact Sciences, Tel-Aviv University.}
$\!\!{}^,$\footnote b {Work supported in part by the US-Israel
Binational Science Foundation and GIF - the German-Israeli Foundation
for Scientific Research.}}
\CERN
\vskip .5truecm
\centerline{and}
\vskip-1.7truecm
\author{S.\ Theisen$^b$}
\address{Sektion Physik, University Munich, Germany}
\vskip-1.2truecm
\abstract{We review the generalization of the work of Seiberg and
Witten on \nex2 supersymmetric $SU(2)$ Yang-Mills theory to $SU(n)$
gauge groups. The quantum moduli spaces of the effective low energy
theory parametrize a special family of hyperelliptic genus $n\!-\!1$
Riemann surfaces. We discuss the massless spectrum and the
monodromies.}
\vskip.4cm
\centerline{{\it Contribution to the Proceedings of the Workshop}}
\centerline{\it ``Physics from the Planck Scale to Electroweak
Scale''}
\centerline{\it Warsaw, September 21 -- 24, 1994}
\centerline{\it and}
\centerline{\it ``28th International Symposium on the Theory of
Elementary
Particles''}
\centerline{\it Wendisch-Rietz, August 29 -- September 3, 1994}
\vfil
\vskip 1.cm
\ni CERN-TH.7538/94\hfill\break
\ni December 1994
\endpage
\baselineskip=14pt plus 2pt minus 1pt
\sequentialequations
\chapter{Introduction}

In two recent papers \doubref\SWa\SWb, Seiberg and Witten have
investigated \nex2 supersymmetric $SU(2)$ gauge theories and solved
for their exact nonperturbative low energy effective action. Some of
their considerations have recently been extended to $SU(n)$ in
\doubref\KLTYa\AF.
We like to review here our work and present
a global describtion of the monodromies of $SU(3)$,
previewing some results to appear in a more complete
account \KLTY.

For arbitrary gauge group $G$, $N=2$ supersymmetric gauge theories
without matter hypermultiplets are characterized by having flat
directions for the Higgs vacuum expectation values, along which the
gauge group is generically broken to the Cartan subalgebra. Thus, the
effective theories contain $r\!=\!{\rm rank}(G)$ abelian \nex2 vector
supermultiplets, which can be decomposed into $r$ \nex1 chiral
multiplets $A^i$ plus $r$ \nex1 $U(1)$ vector multiplets
$W^i_\alpha$. The \nex2 supersymmetry implies that the effective
theory up to two derivatives depends only on
a single holomorphic prepotential $\cF(A)$.
More precisely, the effective lagrangian in \nex1 superspace is
$$
\cL\ =\ {1\over4\pi}{\rm Im}\,\Big[\, \int \!d^4\theta\,\big(\sum
{\del
\cF(A)\over\del A^i}\bar A^i\big) \,+\, \int \!d^2\theta\,{1\over2}
\big(\sum {\del^2 \cF(A)\over\del A^i\del
A^j}W_\alpha^iW_\alpha^j\big)\Big]\ . \eqn\effL
$$
The holomorphic function $\cF$ determines the quantum moduli space
and, in particular, its metric. This space has singularities at
subspaces of complex codimension one, where additional fields become
massless. At these regions the effective action description breaks
down.
A crucial insight is that the electric and magnetic quantum numbers
of the fields that become massless at a given singularity are
determined by the left eigenvectors (with eigenvalues equal to +1)
of the monodromy matrix associated with that singularity.

For $G=SU(2)$ considered in \doubref\SWa\SWb, besides the point at
$u=\infty$ there are singularities at $u=\pm\L^2$, where $\L$ is the
dynamically generated scale of the theory, and $u=\shalf \langle
a^2\rangle$, where $a\equiv A|_{\theta=0}$. On the other hand, $u=0$
is not singular in the exact quantum theory, which means that, in
contrast to the classical theory, no massless non-abelian gauge
bosons arise here (nor at any other point in moduli space). The
singularities at $u=\pm\Lambda^2$ correspond to a massless monopole
and a massless dyon, respectively. The parameter region near
$u=\infty$ describes the semiclassical regime and is governed by the
perturbative beta function with only one-loop
contributions \doubref\SCHIF\NSbeta. It gives rise to a non-trivial
monodromy as well (arising from the logarithm in the effective
coupling constant), but there are no massless states associated with
it.
Although in refs.\doubref\SWa\SWb\ it appears as an assumption, we
believe that it can be proven that only two extra singularities
besides
infinity can exist for  $G=SU(2)$. On physical grounds this would
follow from the fact that to the best of our understanding of gauge
theories only one mass scale $\Lambda$ is generated. Having more
singularities would probably imply that there are more independent
scales in the problem. What is easy to prove is that if indeed there
are only two singularities at finite points, they
correspond up to conjugation to monopole and
dyon states with (magnetic;electric)
charges $(1;-4 n)$ and $(1;-2-4 n)$ ($n\in\ZZ$).
The proof follows by looking for solutions
to the set of diophantine equations $M_{(g_1;q_1)}\,
M_{(g_2;q_2)}=M_\infty$, where
$$
M_{(g;q)}\ =\ \pmatrix{1-gq& - q^2\cr g^2&1+gq} \eqn\sutwomono
$$
is the monodromy matrix for a massless dyon with electric charge
$q$ and magnetic charge $g$ and $M_{\infty}=
\left({-1\atop0}{-4\atop -1}\right)$.
We also have considered the situation of four
singular points and could exclude solutions to the corresponding
diophantine equations, $\prod_{i=1}^4 M_{(g_i;q_i)}=M_\infty$ for
$g_i,q_i\leq 10$.

The singularity structure and knowledge of the monodromies allow to
completely determine the holomorphic prepotential $\cF$. The
monodromy group is $\Gamma_0(4)\subset SL(2,\ZZ)$ consisting of all
unimodular integral matrices $({a\atop c}{b\atop d})$ with $b=0$ mod
4. The matrices act on the vector $(a_D;a)^t$, where $a_D$ is the
magnetic dual of $a$, with $a_D\equiv {\del F(a)\over\del a}$.
The quantum moduli space is the $u$-plane punctured at $\pm \L^2$ and
$\infty$ and can be thought of as $\IH/\Gamma_0(4)$ ($\IH$ is the
upper
half-plane).

The basic idea \SWa\ in solving for the effective theory is to
consider a family of holomorphic curves whose monodromy group
is $\Gamma_0(4)$ and which can be represented as follows:
$$
y^2\ =\ (x^2-u)^2-\L^4 \ .
\eqn\curve
$$
By transforming to Weierstrass form, this curve can be shown to be
equivalent to the curve given in \SWb, which, in contrast to the
curve given in \SWa, is the form appropriate for generalization to
$SU(n),n>2$. The curves \curve\ represent a double cover
of the $x$-plane with the four branch points at
$\pm\sqrt{u\pm\Lambda^2}$, and describe a genus one Riemann surface.
That is, the quantum moduli space of the $SU(2)$ super Yang-Mills
theory coincides with the moduli space of a particular torus; this
torus becomes singular when two branch points in \curve\ coincide.
The derivatives of the electric and magnetic coordinates $(a_D;a)^t$
with respect to $u$ are given by the periods of the holomorphic
one-form ${dx\over y}$ with respect to a symplectic homology basis.
Their ratio, the modular parameter $\tau$, is positive definite
and well-defined in the $u$-plane, and equals the metric of the
moduli space.
Integrating the periods yields $a(u), a_D(u)$ and integrating $a_D$
finally determines the prepotential $\cF(a)$.

We want to indicate next how these ideas of Seiberg and Witten
generalize to gauge groups $G=SU(n)$. To be specific, we will mainly
consider $G=SU(3)$, but the generalization to higher $n$ is
straightforward.

\chapter{Semi-classical Regime}

\ni We will denote the gauge invariant order parameters
(Casimirs) of $SU(n)$ by
$$
u_{k}={1\over k}\Tr\langle\phi^{k}\rangle\,,\quad k=2,\dots, n
\eqn\uvdef
$$
where we can always take the scalar superfield component to be
$\phi=\sum_{k=1}^{n-1}a_k H_k$ with $H_k=E_{k,k}-E_{k+1,k+1},\,
(E_{k,l})_{i,j}=\delta_{ik}\delta_{jl}$. The anomaly free global
$\ZZ_{2n}$ subgroup of $U(1)_{\cal R}$ acts as $u_k\to e^{i\pi k/n}
u_k$. For $SU(3)$ this means that classically $u\equiv
u_2={\a}^2+{\b}^2-\a\b$, $v\equiv u_3=\a\b(\a-\b)$ with $\ZZ_6$
action $u\to e^{2\pi i/3}u$, $v\to -v$. For generic eigenvalues of
$\phi$, the $SU(3)$ gauge symmetry is broken to $U(1)\times U(1)$,
whereas if any two eigenvalues are equal, the unbroken symmetry is
$SU(2)\times U(1)$. These classical symmetry properties are encoded
in the following, gauge and globally $\ZZ_6$ invariant discriminant:
$$
\D_0\ =\ 4 u^3 - 27 v^2\ =\ (\a+\b)^2(2\a-\b)^2(\a-2\b)^2\
\ .\eqn\clasD
$$
The lines $\D_0=0$ in $(u,v)$ space correspond to unbroken
$SU(2)\times U(1)$ and have a cusp singularity at the origin, where
the $SU(3)$ symmetry is restored. As we will see, in the full quantum
theory this cusp is resolved, $\Delta_0\to\Delta_\Lambda=4 u^3-27
v^2+O(\Lambda^6)$, which, in particular, prohibits a phase with
massless non-abelian gluons. Other singularities will however appear,
signalling the appearance of massless monopoles and dyons in the
spectrum.

The prepotential $\cF$ in the semi-classical, perturbative regime can
easily be computed with the result
$$
\cF_{{\rm {semi-\atop class}}}\ =\ {i\over 4\pi}
\sum_{i<j}^3(e_i-e_j)^2\log[(e_i-e_j)^2/\L^2] \ .
\eqn\clasF
$$
Here, $e_i$ denote the roots of the equation
$$
\wan2(x,u,v)\ \equiv\ x^3 - u\,x - v\ =\ 0\ ,
\eqn\Wsing
$$
whose bifurcation set is given by the discriminant $\D_0$ in \clasD,
i.e.
$$
\eqalign{
e_1-e_2\ &=\ \a+\b\cr
e_1-e_3\ &=\ 2\a-\b\cr
e_2-e_3\ &=\ \a-2\b.\cr
}\eqn\eiarel
$$
The Casimirs $u,v$ are gauge invariant and, in particular,
invariant under the Weyl group $\cW$ of $SU(3)$. This group is
generated by any two of the transformations
$$
\eqalign{
r_1:\ \  (\a,\b)\ &\rightarrow\ (\b-\a,\b)\cr
r_2:\ \  (\a,\b)\ &\rightarrow\ (\a,\a-\b)\cr
r_3:\ \  (\a,\b)\ &\rightarrow\ (-\b,-\a)\ .\cr
}\eqn\weyldef
$$
Due to the multi-valuedness of the inverse map $(u,v)\to(\a,\b)$,
closed paths in $(u,v)$ space will, in general, not close in
$(\a,\b)$ space, but will close up to Weyl transformations. Such a
monodromy will be non-trivial if a given path encircles a singularity
in $(u,v)$ space --- in our case, the singularities will be at
``infinity'' and along the lines where the discriminant vanishes.

It is indeed well-known \Arn\ that the monodromy group of the simple
singularity of type $A_2$ \Wsing\ is given by the Weyl group of
$SU(3)$, and acts as Galois group on the $e_i$ (and analogously for
$\wan{n-1}$ related to $SU(n)$). This will be the starting point for
our generalization.

What we are interested in is however not just the monodromy
acting on $(\a,\b)$, but the monodromy acting on $(\ad,\bd;\a,\b)^t$,
where
$$
a_{Di}\ \equiv\ \cF_i\ =\ {\del\over\del a_i}\,\cF(\a,\b)\ .
\eqn\aD
$$
Performing the Weyl reflection $r_1$ on $(\a,\b)^t$, we easily
find
$$
\left(\matrix{\cF_1\cr\cF_2\cr}\right)\ \rightarrow\
\pmatrix{-1 & 0\cr 1 & 1 \cr}\left(\matrix{\cF_1\cr\cF_2\cr}\right)+
N\,\pmatrix{6 & -3\cr -3 &-3\cr}\left(\matrix{\a\cr\b\cr}\right)\ .
\eqn\ronead
$$
The second term, i.e., the ``quantum shift'', arises from the
logarithms and is not determined by the finite, ``classical'' Weyl
transformation acting on the $a_i$,
but rather depends on the particular path in $(u,v)$ space. (We
considered in \ronead\ only paths for which all three logarithms in
\clasF\ contribute with the same sign; other paths do exist where the
logarithms contribute differently and yield a quantum shift matrix
different from that in \ronead\ \KLTY). For example, for the closed
loop given by $(u(a_i(t)),v(a_i(t))$ for $t\in[0,1]$, where
$a_1(t)=e^{i\pi t}a_1 +\shalf(1-e^{i\pi t})a_2$, $a_2(t)=a_2$ we find
$N=1$. Therefore, the matrix representation of $r_1$ acting on
$(\ad,\bd;\a,\b)^t$ is:\foot {In \KLTYa\ we represented these
matrices for rescaled $a_i$. However, the present normalization is
more appropriate if one wants the other monodromies to be given by
integral matrices.}
$$
r_1\ =\ \pmatrix{-1 & 0 & 6 & -3\cr 1 & 1 & -3 & -3\cr 0 & 0 & -1&
1\cr
0 & 0 & 0 & 1 \cr }\ \equiv\ \rc_1\, T^{-3}\ ,
\eqn\ronedef
$$
where $\rc_1$ is the ``classical'' Weyl reflection (given by the
block diagonal part of $r_1$), and $T$ the ``quantum monodromy''
$$
T\ =\ \pmatrix{\textstyle\bfone & C\cr 0&\bfone\cr},\ \ \qquad{\rm
where}\ \ C\ =\ \pmatrix{\textstyle2 & -1\cr-1&2\cr}
\eqn\quantmon
$$
is the Cartan matrix of
$SU(3)$. The other Weyl reflections are given analogously by
$r_i=\rc_i\,T^{-3}$. The $r_i$ are related to each other by
conjugation,
and, in particular, rotate into each other via the Coxeter
element, $\rc_{{\rm cox}}=\rc_1\rc_2$.

\chapter{The Curves for $SU(n)$}

Let us now generalize the $SU(2)$ curve to a sequence of curves
$\cC_n$ whose moduli spaces are supposed to coincide with the quantum
moduli space of effective low energy $N=2$ supersymmetric $SU(n)$
Yang-Mills theories.

We first list the requirements we impose on the curves
$\cC_n$: (i) We seek surfaces with $2(n-1)^2$ periods (corresponding
to $(\partial_{u_j}\adi;\partial_{u_j}\ai)$), whose period matrices
$\Omega_{ij}={\partial a_{Di}\over\partial a_j}$ are positive
definite. (ii) For $\L\to0$ the classical situation must be
recovered. That is, the discriminant of $\cC_n$ should have, for
$\L=0$, a factor of $\D_0$ (cf. eq.\clasD). This means that for
$\L=0$ the curves should have the form
$y^m=\cC_n(x)\equiv\wan{n-1}\!\! \times\!(\dots)$ for some $m$. (iii)
The curves must behave properly under the cyclic global
transformations acting on the Casimirs $u_k$; in other words, there
should be a natural dependence on Casimirs, for all groups. Finally,
from \SWb\ we know that $\L$ should appear in $\cC(x)$ with a power
that corresponds to the charge violation of the one-instanton
process, which is $2n$ for $SU(n)$.

Taking these requirements together leads us to consider the following
genus $g\!=\!n\!-\!1$ hyperelliptic curves for $SU(n)$:
$$
y^2\ =\ \cC_n(x)\ \equiv\ \Big(\wan{n-1}(x,u_i)\Big)^2 - \L^{2n}\ ,
\eqn\theCurve
$$
where
$$
\wan{n-1}(x,u_i)\ =\ x^n - \sum_{i=2}^n u_i\,x^{n-i}.
\eqn\LGpots
$$
are the $A$-type simple singularities related to $SU(n)$.
The square of $W$ reflects having both electric and
magnetic degrees of freedom. Since there is a general relationship
\Arn\ between Arnold's simple A-D-E singularities, perturbations by
Casimirs, monodromy and Weyl groups, we conjecture that \theCurve\
describes surfaces for the other simply laced gauge groups $G$ as
well. One simply replaces $\wan{n-1}(x,u_i)$ by the corresponding D-
or E-type singularity, and $\L^{2n}$ by $\L^{2h}$, where $h$ is the
corresponding Coxeter number.

\ni For the following it will be useful to write
$$
\eqalign{
\cC_n(x)\ &=\
\Big(\wan{n-1}(x,u_i)+\L^n\Big)\Big(\wan{n-1}(x,u_i)-\L^n\Big)\cr
&=\ \prod_{i=1}^n(x-e_i^+)(x-e_i^-)\ .\cr}
\eqn\factcurve
$$
Critical surfaces occur whenever two roots of $\cC_n(x)$ coincide,
that
is, whenever the discriminant $\D_\L=\prod_{i<j}(e_i^\pm-e_j^\pm)^2$
vanishes. We expect this to happen when monopoles or dyons,
whose quantum numbers are determined by the corresponding
monodromy matrices, become massless.
For example, for $G=SU(3)$ the quantum
discriminant is:
$$
\D_\L\ =\ \L^{18}\D_\L^+\D_\L^-\ ,  \ \ \
\ \ \D^\pm_\L\ =\ 4 u^3-27(v\pm\L^3)^2.
\eqn\thefullD
$$
By construction, the hyperelliptic curves \theCurve\ are represented
by branched covers over the $x$-plane. More precisely, we have $n$
$\ZZ_2$ cuts, each linking a pair of roots $e^+_i$ and $e^-_i$,
$i=1,\dots,n$. As an example, we present the picture for $G=SU(3)$ in
\lfig\figone. In the classical theory, where $\L\to0$, the branch
cuts shrink to $n$ doubly degenerate points: $e^-_i\to e^+_i\equiv
e_i$. These points, given for $SU(3)$ in eq.\ \eiarel, correspond to
the weights of the $n$-dimensional fundamental representation (the
picture represents a deformed projection of the weights onto the
unique Coxeter eigenspace with $\ZZ_n$ action). This means that the
branched $x$-plane transforms naturally under the finite
``classical'' Weyl group that permutes the points. This finite Weyl
group is just the usual monodromy group of the $A_{n-1}$ singularity
alluded to earlier. In the quantum theory, where the degenerate
points are resolved into branch cuts, there are in addition
possibilities for ``quantum monodromy'', which involves braiding of
the cuts.
\figinsert\figone{Branched $x$-plane with cuts
linking pairs of roots of $\cC_3=0$. We depicted a choice of basis
for the homology cycles that is adapted to the cyclic $\ZZ_3$
symmetry
generated by the classical Coxeter element. Note that the locations
of the cuts are $\ZZ_3$-symmetric only for $u=0$.
}{2.0in}{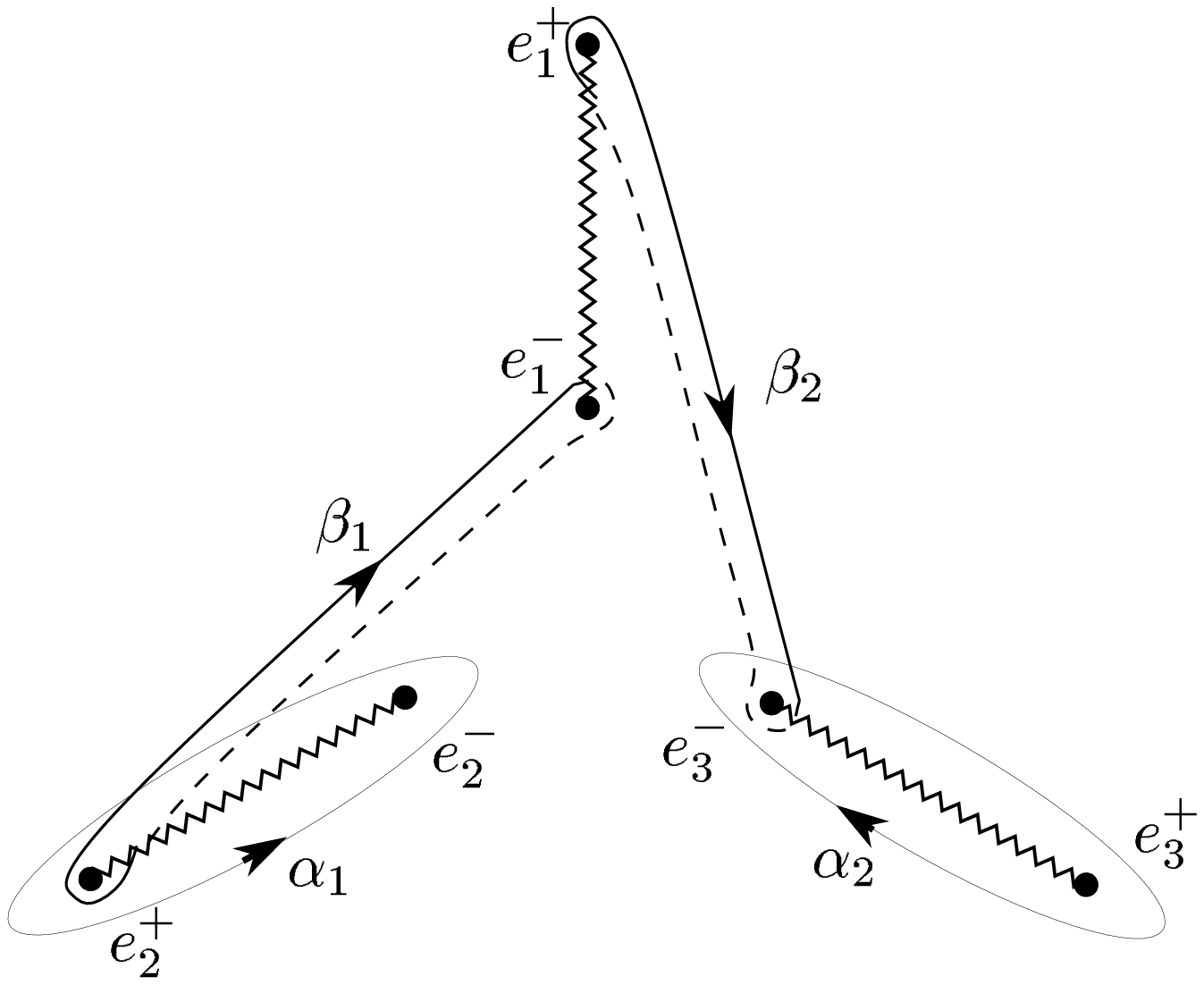}

\chapter{Monodromies for G=SU(3)}

The semi-classical monodromy discussed so far was
 determined from properties
relying on asymptotic freedom of the $SU(3)$ theory. We now turn to
the monodromies $M_{\lambda}$ around singularities related to
massless dyons of charges $\lambda\equiv(\vec g,\vec q)^t\equiv
(g_1,g_2;q_1,q_2)^t$ (the
charge vectors will be given in a basis generated by the two
fundamental weights of $SU(3)$). Analogous to \SWa, the charge
vectors are left eigenvectors of the corresponding monodromy
matrices with eigenvalue equal to one, i.e.
$\lambda^t{M}_{\lambda}=\lambda^t$.
Obviously, the monodromies of any two dyons whose charge
vectors are related via $\lambda_2=N^t\lambda_1$, ${N}\in
Sp(4,\ZZ)$, are conjugated:
${M}_{\lambda_2}={N}^{-1}{M}_{\lambda_1}{N}$. If we require the
monodromies to depend only on the charges of the corresponding
states, the matrices are determined to take the following form,
which is the generalization of \sutwomono:
$$
{M}_{(\vec g; \vec q)}=
\pmatrix{\bfone - \vec q\otimes \vec g& -\vec q\otimes \vec q\cr
\vec g\otimes \vec g&\bfone +\vec g\otimes \vec q}
\eqn\monmatrix
$$
This form of the matrix is obviously the same for all $SU(n)$.
Here we have fixed some freedom with hindsight to our results below.
Note that for purely magnetically charged monopoles, where $q_i=0$,
one
has: $a_{Di}\to a_{Di}$, and this reflects the fact that the $a_{Di}$
are good local coordinates in the vicinity of the singular loci where
the monopoles become massless.

We now indicate how to determine the monodromies $M_{\lambda}$ of the
curves $\cC_n$. Fixing a base point
$u_0$ in the moduli space
${\cal M}_n$ of $\cC_n$, there is a homomorphism of the fundamental
group
$\pi_1({\cal M}_n,p)$ into $Sp(2n\!-\!2;\ZZ)$ whose image
is the monodromy
group. Note that the singular locus of $\cC_n$, $\Delta_\Lambda=0$,
does itself have singularities. For $n=3$ there are cusps at $u=0,\,
v=\pm\Lambda^3$ ($4\delta u^3-27\delta v^2=0$) and nodes at
$u^3={27\over4}\Lambda^6,\, v=0$ ($\rho\,\delta u^2-\delta
v^2=0,\,\rho^3=\Lambda^6/4$), as discussed in \AF. We have depicted
in \lfig\figtwo\ the moduli space for real $v$.
\figinsert\figtwo{$SU(3)$ quantum moduli space for real $v$. The six
lines are the singular loci where $\Delta_\Lambda=0$ and where
certain dyons become massless. At each of the three crossing points
at $v=0$, two mutually local dyons become simultaneously
massless, and the theory is semi-classical in the corresponding dual
variables. The various markings of the lines indicate how the
association with particular monodromy matrices changes when moving
through the cusps. }{4.in}{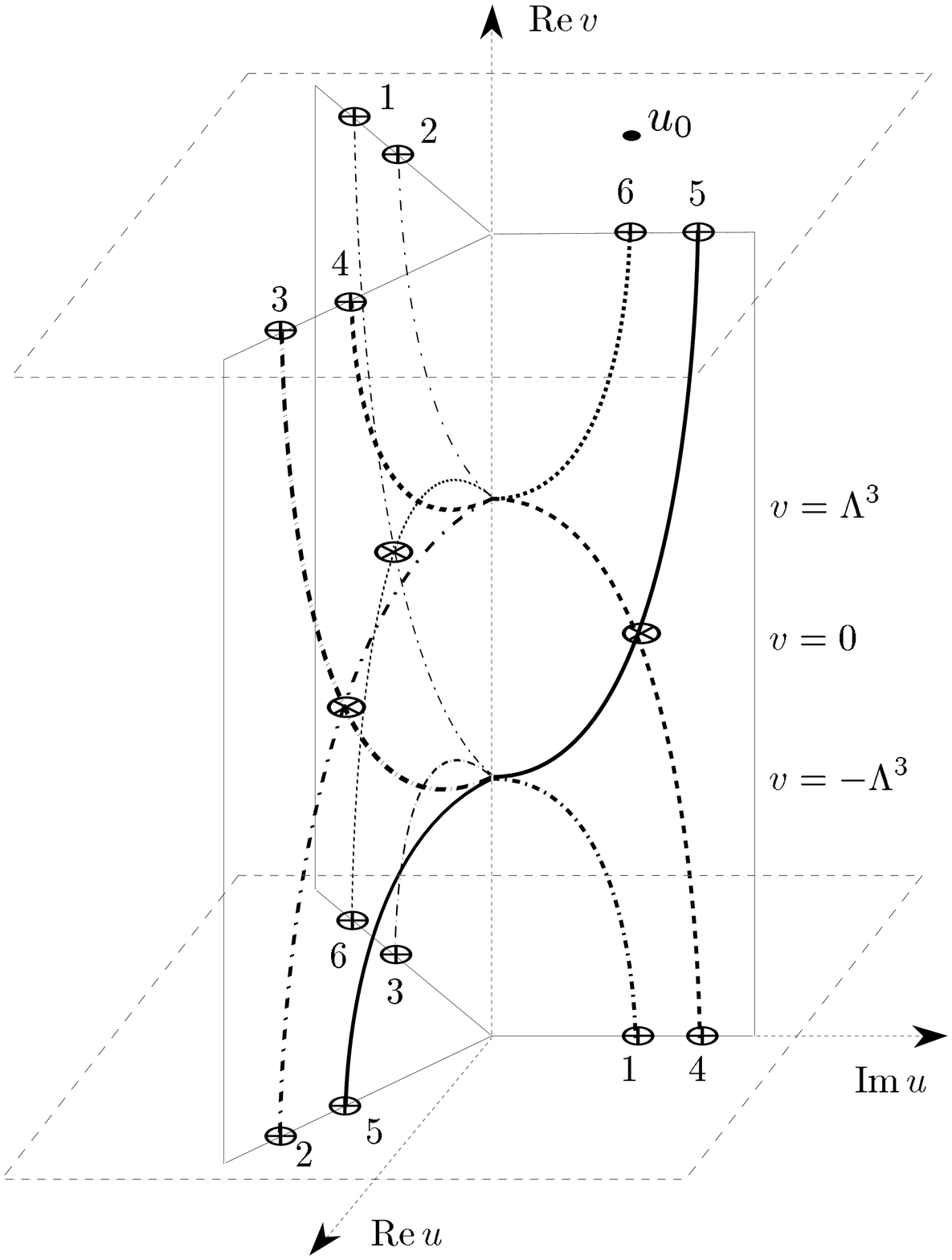}

The monodromy matrices reflect the action of braiding
and permuting the cuts on the vector $(\adi;\ai)^t$. This action is
expressed in terms of the action on the homology cycles via
$$
\adi\ =\ \oint_{\beta_i}\lambda\ ,\ \ \ \ \ \
\ai\ =\ \oint_{\alpha_i}\lambda\ ,
\eqn\aints
$$
where $\alpha_i,\beta_j$ is some symplectic basis of
$H_1(\cC_n;\ZZ)$:
$\langle\alpha_i,\beta_j\rangle=-\langle\beta_j,\alpha_i \rangle=
\delta_{ij}$, $\langle\alpha_i, \alpha_j\rangle=\langle\beta_i,
\beta_j\rangle=0$, $i,j=1,\ldots,g$ and $\lambda$ is some suitable
chosen meromorphic differential.
{}From the theory of Riemann
surfaces it is clear that the monodromy group is contained in
$Sp(2g,\ZZ)=Sp(2n\!-\!2,\ZZ)$. For $G=SU(3)$, we have depicted our
choice of homology basis in \lfig\figone.

The holomorphic differentials on a genus $g=n-1$
hyperelliptic curve are $\omega_{n-i}=
x^i{dx\over y}$, $i=0,\ldots,g-1$. The $g\times 2g$ period matrix
is $(A,B)=A({\bf1},\Omega)$ with $A_{ij}=\int_{\alpha_j}\!\omega_i$
and $B_{ij}=\int_{\beta_j}\!\omega_i$, which are related to
$(\adi,\ai)$ as follows:
$$
A_{ij}={\partial a^j(u)\over \partial u_i},
\quad B_{ij}={\partial a_{D}^j(u)\over \partial u_i}\ .
\eqn\Ident
$$
Note that due to \Ident, the second Riemann bilinear relation,
${\rm Im}\,\Omega>0$, ensures the positivity of the metric
$$
(ds)^2={\rm Im}\,{\partial{\cal F}\over\partial a_i\partial a_j}
da_i d\bar a_j={\rm Im}\sum_{i=1}^g
da_{D}^i d\bar a_i\,.
$$
This generalizes the situation at genus one \SWa.
Note also that \Ident\ represents a non-trivial integrability
condition
for the periods, $\del_iA_{jk}=\del_jA_{ik}$,
$\del_i B_{ik}=\del_j B_{ik}$,
which holds for our parametrization of the curve \theCurve.
The periods, which can be obtained as solutions of the Picard-Fuchs
equations \KLTY, can thus be integrated
to yield $a_{D\, i}(u,v)$, $a_i(u,v)$ and by further integration
to yield ${\cal F}(u,v)$. Alternatively one can choose the
meromorphic
differential $\lambda$ in \aints\ s.t. $\omega_i \sim\del_i \lambda$
up to exact forms, e.g. $\lambda\sim(3 x^3-u x){dx\over y}$ as in
\AF, and try to integrate \aints.

Now, if $\nu$ denotes the vanishing cycle for the braiding of the
branch points induced by a loop $\gamma$ in moduli space, the
action on the homology cycle $\delta\in H_1({\cal C}_n;\ZZ)$
can be simply obtained via the
Picard-Lefshetz
formula \Arn
$$S_\nu\delta=\delta+
\langle\delta,\nu\rangle\nu
$$
We find that when decomposing the given
vanishing cycle as $\nu=\sum_{i=1,2}(q_i\alpha_i+g_i\beta_i)$, the
action on the homology basis $(\beta_1,\beta_2,\alpha_1,\alpha_2)^t$
is given by a monodromy matrix $M_\lambda$ precisely of the form
\monmatrix, with labels just given by the expansion coefficients:
$M_\lambda\equiv M_{(g_1,g_2;q_1,q_2)}$. That is, we can read off the
electric and magnetic quantum numbers of a given massless dyon just
by looking at the picture of the corresponding vanishing cycle !

More specifically, we studied the monodromies of $G=SU(3)$ by fixing
a base point $u_0$ and by carefully tracing the effects of loops in
moduli space on the motions of the branch points in the $x$-plane.
With reference to results by Zariski and van Kampen (cf.\ \CDFLLR\
and references therein) it suffices to study loops in a generic
complex line through the base point. Across cusps and nodes the
monodromies are related through the ``van Kampen relations''. In
\lfig\figthree\ the six marked points are the intersections of the
complex $u$-plane at ${\rm Re}(v)={\rm const}>\Lambda^3,\
{\rm Im(}v)=0$ with the singular set $\Delta_\Lambda=0$.
The monodromies around the six loops
$\g_i$ can be characterized by the corresponding vanishing cycles,
which are depicted in \lfig\figfour. According to what we said above,
the quantum numbers of the dyons that become massless at the
encircled singular lines in moduli space can be directly obtained
from \lfig\figfour, by comparing the vanishing cycles with the basis
cycles in \lfig\figone. The corresponding monodromy matrices
\monmatrix\ then turn out to be:
$$
\eqalign{
M_{\lambda_1}\ =\ M_{(1,1;-1,0)},\quad&M_{\lambda_2}\ =\
M_{(1,1;0,1)},\cr
M_{\lambda_3}\ =\ M_{(1,0;-1,1)},\quad&M_{\lambda_4}\ =\
M_{(1,0;1,0)},\cr
M_{\lambda_5}\ =\ M_{(0,1;0,-1)},\quad&M_{\lambda_6}\ =\
M_{(0,1;-1,1)}\ .}
\eqn\sixmonodroms
$$

\figinsert\figthree{Loops $\g_i$ in the $u$-plane at ${\rm
Re}(v)=const>\Lambda^3,\ {\rm Im(}v)=0$ (cf., \lfig\figtwo). $u_0$
denotes
the
base point. A loop encircling all the points
yields essentially the monodromy matrix $r_1$ given in
\ronedef.}{2.0in}{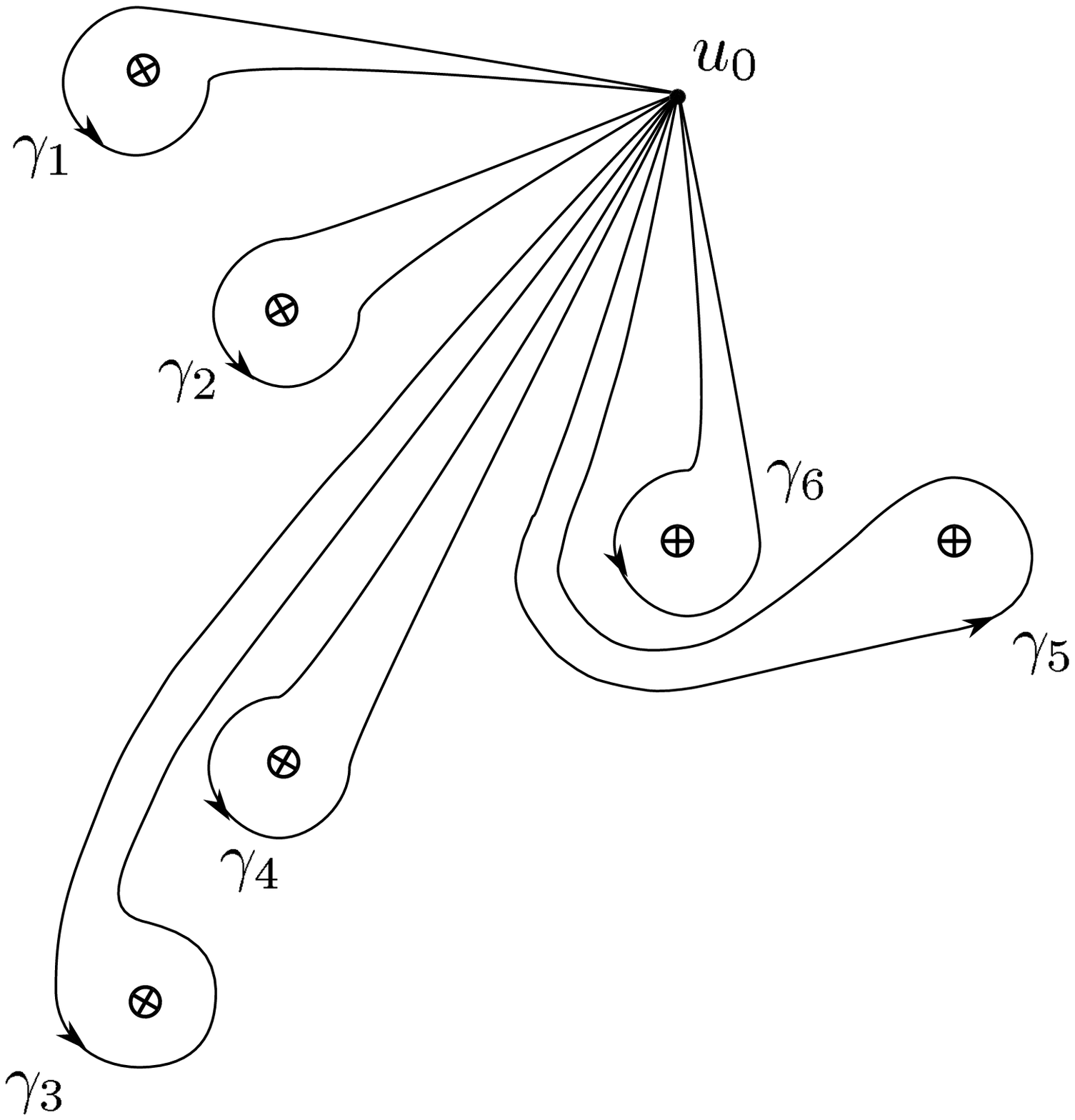}

\figinsert\figfour{Vanishing cycles $\nu_i$ associated with the loops
$\g_i$ in \lfig\figthree. The dyon charges can be directly inferred
from this picture, by expanding the cycles in terms of the homology
basis given in \lfig\figone\ (e.g. $\nu_6=\beta_2-\alpha_1+\alpha_2$
$\Rightarrow$ $\lambda_6=(0,1;-1,1)$ ). We have depicted here only
the
paths on the
upper sheet, and not the return paths on the lower
sheet.}{3.0in}{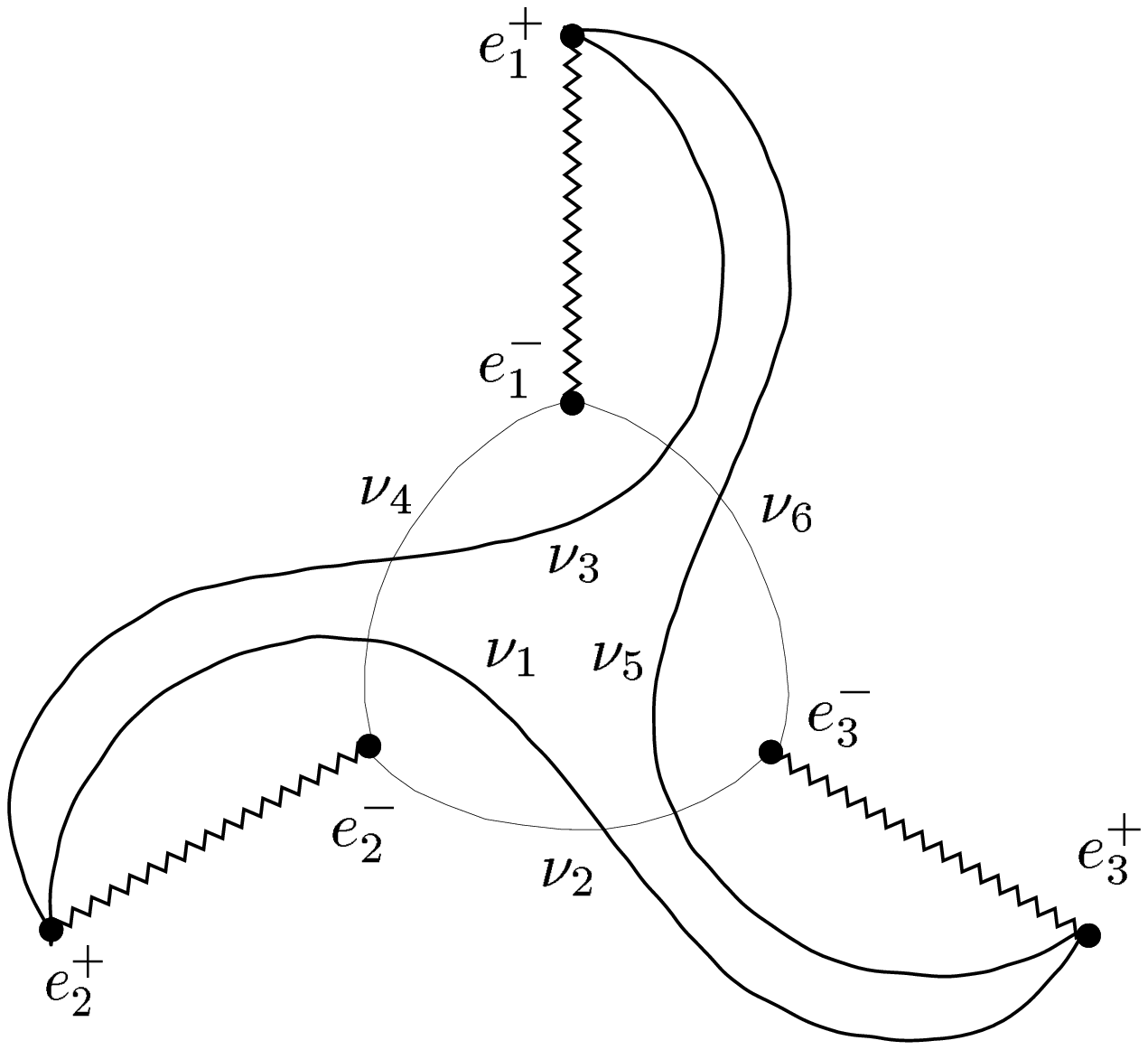}

We remark that the charge vectors of each pair of lines which cross
in the nodes at $u=0$ satisfy $\lambda_i^t\,\Sigma\, \lambda_j=0$,
where $\Sigma=\left({0\atop{\bf -1}}{{\bf 1}\atop0}\right)$ is the
symplectic metric \AF. This is a necessary condition for two types of
dyons to condense simultaneously \thooft. (It follows from
\monmatrix\ that if two charge vectors are symplectically orthogonal,
the corresponding monodromy matrices commute.) More precisely, the
charge labels $[\lambda_i,\lambda_j]$ of the condensing pairs are
$\big\{[\lambda_1,\lambda_6],\,[\lambda_2,\lambda_3],\,
[\lambda_4,\lambda_5]\big\}$. Note that by appropriately
changing the homology basis, one can always
conjugate the charge vectors of any given one such pair of dyons to
purely magnetic charges, ie., to $(1,0;0,0)$ and $(0,1;0,0)$.
At the nodes both dual
$U(1)$'s are weakly coupled and one can verify that the monodromies
are consistent with the beta functions of the effective dual theory
that contains two monopole hypermultiplets.

One type of monodromy at infinity is given by a loop encircling all
the singular points in \lfig\figthree, that is, by the product of the
matrices in \sixmonodroms. It turns out to be precisely the monodromy
\ronedef\ deduced from the semi-classical effective action \clasF, up
to a change of basis:
$$
M_\infty \equiv \prod_{i=1}^6\,M_{\lambda_i}\ =\
S^{-1}\,(r_1)^{-1}S,
$$
with $S=\left({{\bf 1}\atop0}{s\atop{\bf 1}}\right)$,
$s\equiv\left({-2\atop0}{0\atop2}\right)$. (In this basis, the other
matrices $r_2$, $r_3$ can be obtained as well, by starting from
a different base point.) We take this as a non-trivial physical
consistency check.

One can also consider monodromies for paths lying in other complex
lines in the moduli space, like in the line $u=const.$ as
discussed in \AF. This plane is however not in generic position but
is special, in that it does not cut through all six, but only through
four singular lines. (The corresponding monodromy matrices form a
subset of
the matrices in \sixmonodroms). The total monodromy obtained by
encircling the four points in the $v$-plane is different from the
monodromy in the $u$-plane discussed above, however it is closely
related to it: its classical (block-diagonal) part is just given by
the
Coxeter element $\rc_{{\rm cox}}\equiv\rc_1\rc_2$ of the Weyl group
of $SU(3)$. The Coxeter element is of order three, and its $\ZZ_3$
action corresponds to the cyclic rotation of the cuts in
\lfig\figone. The complete monodromy in the $v$-plane, which includes
the quantum shift, is the same as the semi-classical
monodromy obtained in \AF, up to change of basis. It can be obtained
from the effective action \clasF\ as well, by chosing an appropriate
path.

This Coxeter monodromy is closely related to a well-known fact in
singularity theory \Arn. Here one considers perturbations
$W=W_0(u_k=0)+\Lambda$, where the parameter circles around the
origin: $\Lambda=e^{2\pi i t}$, $t\in[0,1]$. This induces what is
called, ironically, ``classical monodromy'', and for the A-D-E simple
singularities, this classical monodromy is precisely given by the
Coxeter element of the corresponding A-D-E type Weyl group. For our
curves \theCurve, such loops in the $\Lambda$-plane indeed reproduce
the above-mentioned Coxeter monodromy in the $v$-plane, if we set
$u,v=0$. On the other hand, if $|\Lambda|<|v|$, these loops do not
induce the cyclic rotation of the cuts in \lfig\figone, but only
induce simultaneous braiding of the cuts. This corresponds to pure
``quantum monodromy'', given by the shift
matrix $T$ in \quantmon\ \KLTYa.

Summarizing, the classical piece of the total monodromy in the
$u$-plane is given by any one of the Weyl group generators, say
$\rc_1$, whereas for the $v$-plane one obtains the Coxeter element,
$\rc_1\rc_2$. We expect for $SU(n)$ that when looping around infinity
in a parameter plane $u_k$ related to a Casimir of degree $k$, the
classical part of the monodromy is given by a Weyl group conjugacy
class of the corresponding order. In particular, encircling the top
Casimir
parameter plane $u_n$ will induce the Coxeter monodromy, which has
order $n$.

\refout
\end